\documentclass[%
 reprint,
 amsmath,amssymb,
 aps,
]{revtex4-2}

\usepackage{graphicx}
\usepackage{dcolumn}
\usepackage{bm}
\usepackage{hyperref}
\usepackage{cleveref}
\usepackage{algorithm}
\usepackage{algpseudocode}
\usepackage{graphicx}
\usepackage{comment}

\begin{document}

\preprint{Eigenvector dreaming}

\title{Eigenvector dreaming}
\date{\today}

\author{Marco Benedetti}
\affiliation{
 Dipartimento di Fisica, Sapienza Università di Roma, P.le A. Moro 2, 00185 Roma, Italy
}
\author{Louis Carillo}
\affiliation{Département de physique, ENS Paris-Scalay, Gif-sur-Yvette, France, and Dipartimento di Fisica, Sapienza Università di Roma, P.le A. Moro 2, 00185 Roma, Italy
}
\author{Enzo Marinari}
\affiliation{
 Dipartimento di Fisica, Sapienza Università di Roma, and CNR Nanotec, Roma, and INFN Sezione di Roma, P.le A. Moro 2, 00185 Roma, Italy
}
\author{Marc M\'ezard}
\affiliation{Department of Computing Sciences, Bocconi University, Milano, Italy
}

\begin{abstract}
Among the performance-enhancing procedures for Hopfield-type networks that implement associative memory, Hebbian Unlearning (or dreaming) strikes for its simplicity and its clear biological interpretation. Yet, it does not easily lend itself to a clear analytical understanding. Here we show how Hebbian Unlearning can be effectively described in terms of a simple evolution of the spectrum and the eigenvectors of the coupling matrix. We use these ideas to design new dreaming algorithms that are effective from a computational point of view, and are analytically far more transparent than the original scheme.
\end{abstract}

\maketitle
\section{Introduction}
\label{sec:Intro}
Consider a fully connected network of $N$ binary variables $\{S_i = \pm 1\}$, $i\,\in [1,..,N]$, linked by couplings $J_{ij}$.  The network is endowed with a dynamics 
\begin{equation}
    \label{eq:dynamics}
    S_i(t+1) = \text{sign}\left(\sum_{j=1}^N J_{ij}S_j(t)\right), \hspace{0.7cm} i = 1,..,N
\end{equation}
 which can be run either in parallel (i.e. \textit{synchronously}) or in series (i.e. \textit{asynchronously} in a predetermined or in a random order) over the $i$ indices. This kind of network can be used as an associative memory device, namely for reconstructing an extensive number $P=\alpha N$ of binary patterns $\{\xi^{\mu}_i\}=\pm 1$, $\mu\,\in [1,...,P]$, called \textit{memories}. In this work, we will focus on i.i.d. memories, generated with a probability $P(\xi^{\mu}_i=\pm 1)= 1/2$. We consider a recognition process based on initializing the network dynamics to a configuration similar enough to one of the memories, and iterating \cref{eq:dynamics} asynchronously until a fixed point is reached. The network performs well if such asymptotic states are similar enough to the memories. Whether this is the case depends on the number of patterns one wants to store and on the choice of the coupling matrix $J$. Hebb's learning prescription \cite{hebbOrganizationBehaviorNew1950}  
\begin{equation}
    \label{eq:hebbs_learning_rule}
    J_{ij}^{H} = \frac{1}{N}\sum_{\mu = 1}^p\xi_i^{\mu}\xi_j^{\mu}\,, \qquad J^H_{ii}=0
\end{equation}
used in the seminal work of Hopfield \cite{hopfieldNeuralNetworksPhysical1982}, allows retrieving memories up to a critical capacity $\alpha_c^H\sim0.14$ \cite{amitStatisticalMechanicsNeural1987}. 

In this model even when $\alpha < \alpha_c^H$ memories are not perfectly recalled, but the state of the system always presents a small finite fraction of misaligned spins. This feature is linked to the value of the minimum stability $\Delta_{\text{min}}$, defined as
\begin{equation}
    \label{eq:stabmin}
    \Delta_{\text{min}} \equiv  \text{min}_{i,\mu}\{\Delta_i^{\mu}\},
\end{equation}
where the $\textit{stability}$ $\Delta_i^{\mu}$ is defined by
\begin{equation}
    \label{eq:stab}
    \small
    \Delta_i^{\mu} = \frac{\xi_i^{\mu}}{\sqrt{N}\sigma_i}\sum_{j = 1}J_{ij}\xi_j^{\mu}, \qquad \sigma_i = \sqrt{\sum_{j=1}^N J_{ij}^2/N}.
\end{equation}
The value of the stability tells us if a given pattern is aligned or not to its memory field.  As soon as $\Delta_{\text{min}} > 0$, memories themselves become fixed points of the dynamics \cite{gardnerPhaseSpaceInteractions1989}, allowing error-less retrieval when the dynamics is initialized close enough to one of them.

Several techniques have been developed to build better performing coupling matrices, i.e. to reduce the retrieval error and increase the critical capacity as well as the size of the basins of attraction to which the memories belong \cite{gardnerSpaceInteractionsNeural1988, gardnerTrainingNoiseStorage1989, wongOptimallyAdaptedAttractor1990, dotsenkoStatisticalMechanicsHopfieldlike1991, nokuraUnlearningParamagneticPhase1996}. One such technique is Hebbian Unlearning.

\section{Hebbian Unlearning (HU)}
Inspired by the brain functioning during REM sleep \cite{crickFunctionDreamSleep1983}, the unlearning algorithm \cite{hopfieldUnlearningHasStabilizing1983,vanhemmenIncreasingEfficiencyNeural1990,vanhemmenHebbianLearningIts1998,benedettiSupervisedPerceptronLearning2022} is a training procedure for the coupling matrix $J$, leading to error-less retrieval and increased critical capacity in a symmetric neural network. The coupling matrix is built according to the following iterative procedure:

\begin{minipage}[t]{0.48\textwidth}
    \begin{algorithm}[H]
        \caption{Hebbian unlearning}
        \label{alg:CD}
        \begin{algorithmic}
            \State Initialize $J$ using Hebb's rule \cref{eq:hebbs_learning_rule}
            \For{$d = 1$ to $D_{max}$}
                \State Initialize network to a random state $\sigma$.
                \State Follow dynamics \cref{eq:dynamics} to a stable point $\sigma^*$.
                \For{$i \neq j$}
                    \State $J_{ij} \gets J_{ij} - \frac{\epsilon}{N} \sigma_i^* \sigma_j^*$ 
                \EndFor
            \EndFor
        \end{algorithmic}
    \end{algorithm}
\end{minipage}
\vspace{0.3cm}

\begin{figure}[t]
\centering
\includegraphics[width=1\linewidth]{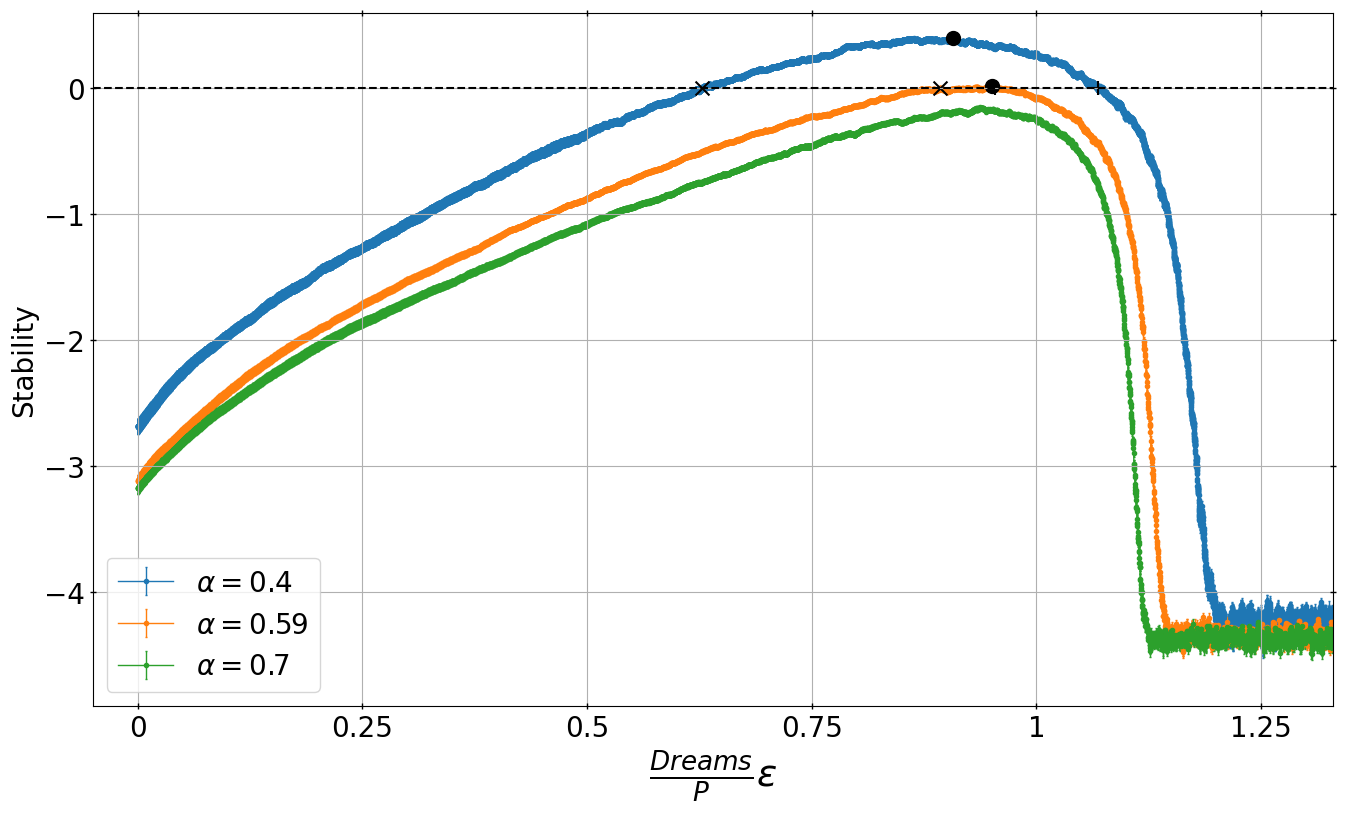}
\caption{The minimum stability $\Delta_{\text{min}}$ as a function of the normalized number of dreams, for different values of $\alpha$. The threshold $\Delta = 0$ is indicated with the gray dotted line. 
For $\alpha<0.59$, $\Delta_{min}$ crosses zero at $D_{in}$, peaks at $D = D_{top}$ and then becomes negative again at $D = D_{fin}$. Where appropriate the three relevant amounts of dreams are indicated: $D = D_{in}$ by "$x$", $D = D_{top}$ by a dot, $D = D_{fin}$ by a "$+$". All measurements are averaged over 50 realizations of the network. $N = 400$, $\epsilon = 10^{-2}$.}
\label{fig:unl_stabmin}
\end{figure}

The learning rate $\epsilon$ and the number of dreams $D_{max}$ are free parameters of the algorithm. The \cref{alg:CD} does not change the diagonal elements of the coupling matrix, which are fixed to $J_{ii}=0$. For sufficiently small values of the learning rate, below the critical load $\alpha<\alpha_c^{HU}\sim 0.6$ the evolution of $\Delta_{min}$ follows a non-monotonic curve as a function of $D_{max}$, as illustrated in \cref{fig:unl_stabmin}. The number of dreams $D = D_{in}$ marks the point where $\Delta_{\text{min}}$ crosses $0$. Here all the memories are fixed points of the dynamics. Other two points, $D = (D_{top},D_{fin})$ are shown in the plot, corresponding to the maximum of $\Delta_{min}$ and the point where $\Delta_{min}$ becomes negative again. The scaling of $(D_{in}, D_{top}, D_{fin})$ was studied in \cite{benedettiSupervisedPerceptronLearning2022}.

\begin{figure}[t]
\centering
\includegraphics[width=1\linewidth]{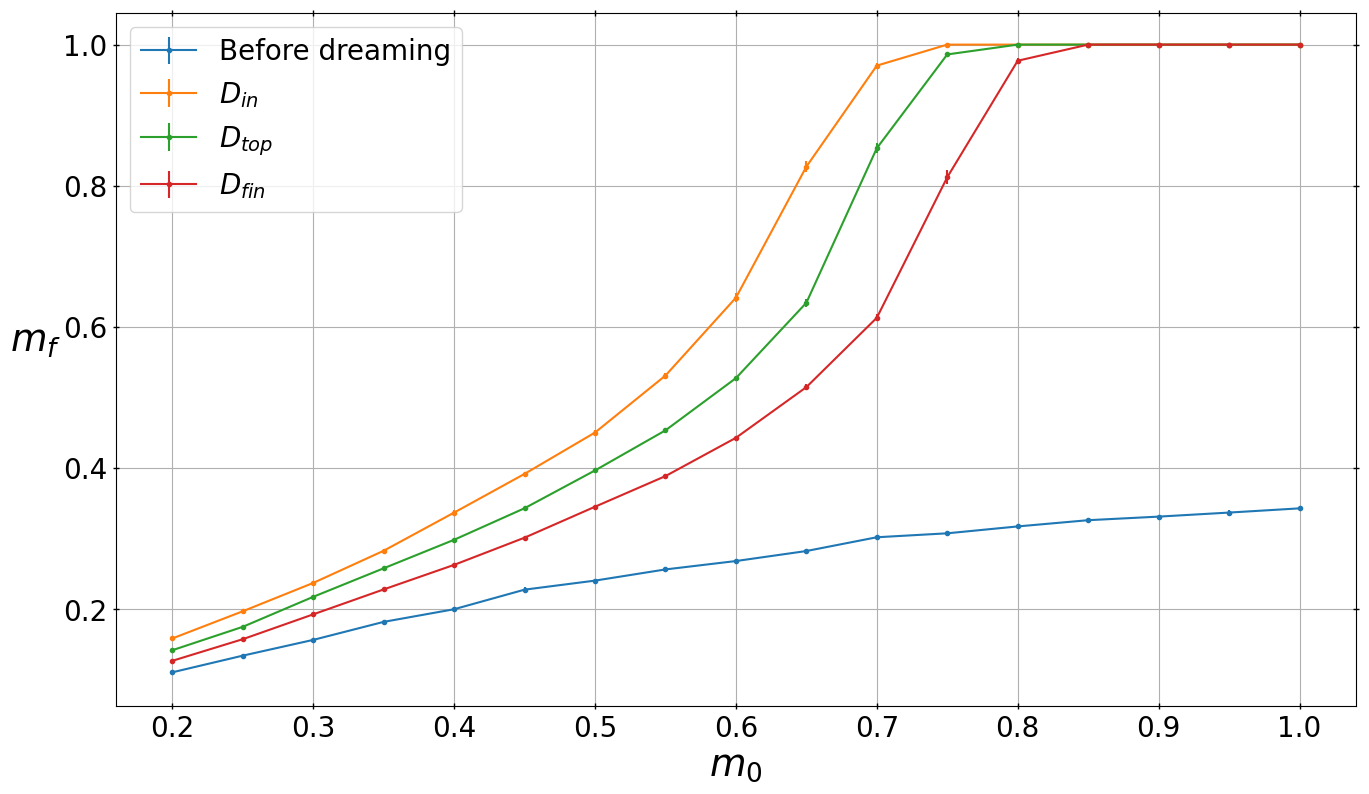}
\caption{Retrieval map $m_f(m_0)$ for the unlearning algorithm at the three relevant steps indicated in Fig. \ref{fig:unl_stabmin}, and before unlearning. All measurements are averaged over 10 realizations of the network. $N = 1000$, $\alpha = 0.4$, $\epsilon = 10^{-2}$. The performance of the algorithm is maximal ad $D=D_{in}$.}
\label{fig:unl_mfvsmi}
\end{figure}

In addition to error-less retrieval,  when $\alpha < \alpha_c^{HU}$, dreaming creates large basins of attraction around the memories.  This can be measured in terms of the \textit{retrieval map}
\begin{equation}
\label{eq:m}
m_f(m_0) \equiv \overline{\Big\langle\frac{1}{N}\sum_{i = 1}^N \xi_i^\mu S_i^{\mu}(\infty)\Big\rangle}\;.
\end{equation}
Here, $\vec{S}^{\mu}(\infty)$ is the stable fixed point reached when the dynamics is initialized to a configuration $\vec{S}^\mu(0)$ having overlap $m_0$ with a given memory $\vec{\xi}^\mu$. The symbol $\overline{\hspace{0.1cm}\cdot\hspace{0.1cm}}$ denotes the average over different realizations of the memories and $\langle \cdot \rangle$ the average over different realizations of $\vec{S}^\mu(0)$. We show in Fig. \ref{fig:unl_mfvsmi} the retrieval map for $N = 1000$ and $\alpha = 0.4$. The performance of HU is best at $D = D_{in}$. Interestingly, as discussed in \cite{benedettiSupervisedPerceptronLearning2022}, the curve relative to Gardner's optimal symmetric perceptron \cite{gardnerSpaceInteractionsNeural1988, gardnerPhaseSpaceInteractions1989} and to unlearning at $D = D_{in}$ coincide with good accuracy. 

\section{Two novel dreaming algorithms}
\label{sec:Two_novel_dreaming_algorithms}
\begin{figure}[t]
\centering
\includegraphics[width=1\linewidth]{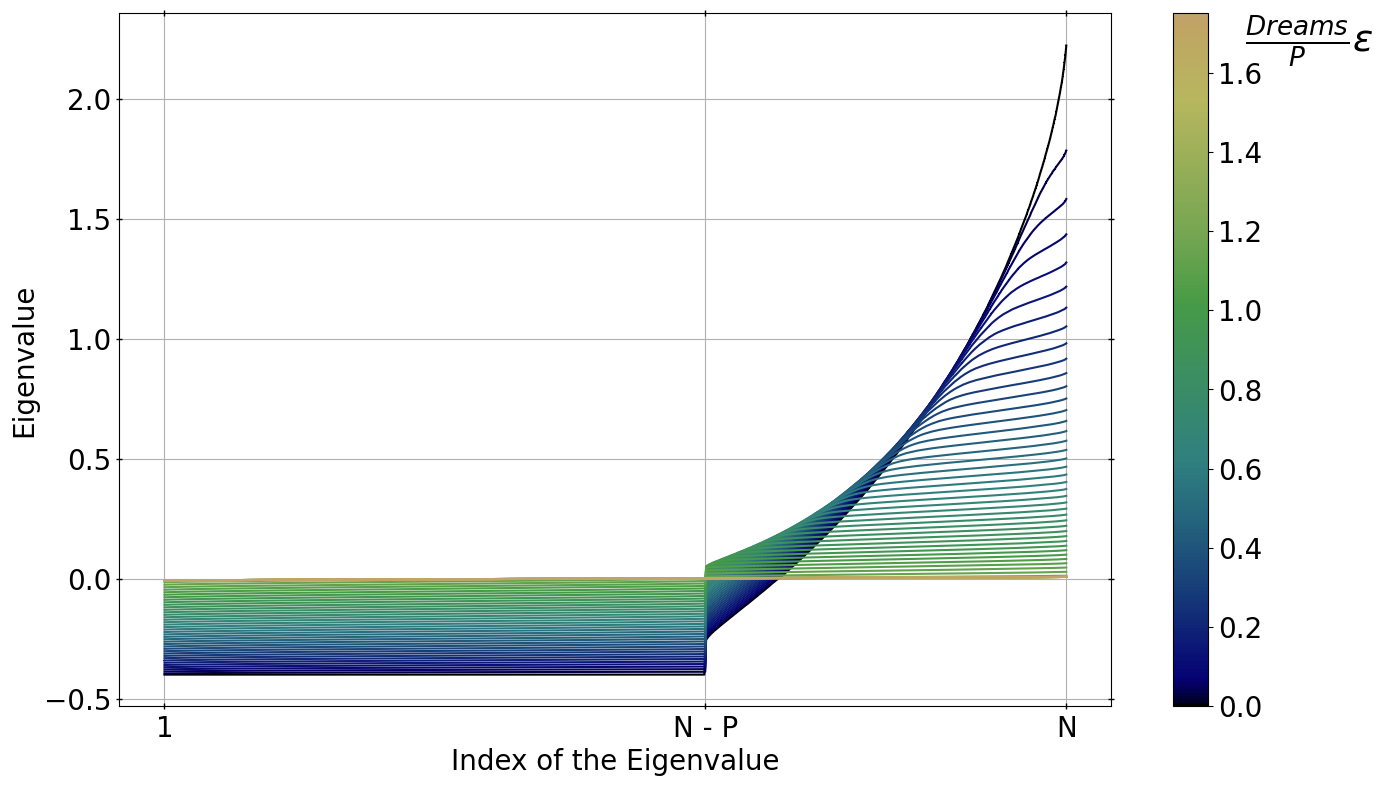}
\caption{On the y-axes, the value of the eigenvalues; on the x-axes their ranking. Curves of different colors correspond to measures of the ranked spectrum taken after different amounts of dreams. Before dreaming, the spectrum is of the Marchenko–Pastur type. HU progressively flattens the high portion of the ranked spectrum}
\label{fig:HU_ranked_spectrum}
\end{figure}

\begin{figure}[t]
\centering
\includegraphics[width=1\linewidth]{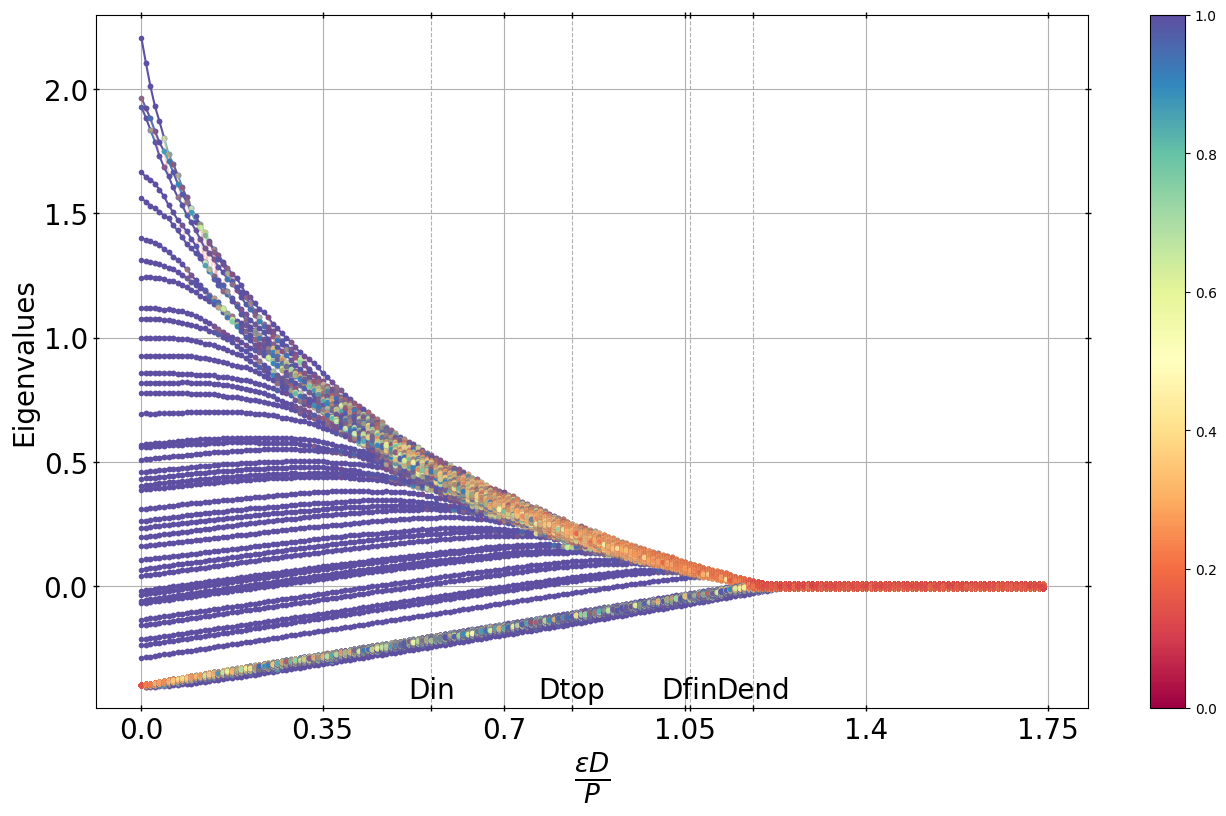}
\caption{On the x-axis, normalized number of steps of the dreaming algorithm. On the y-axes, eigenvalues of the coupling matrix, for one sample, $N=100$. Eigenvalues at different steps of the algorithm are connected by colored lines. Darker colors indicate a high overlap between the corresponding eigenvectors. Only lines corresponding to overlaps larger than $0.1$ are shown. The overlap among subsequent eigenvectors is high, except for the highest and lowest part of the ranked spectrum, where the eigenvalues are effectively degenerate.
\label{fig:HU_panini} 
}
\end{figure}

An interesting interpretation of the HU algorithm emerges when analyzing the evolution of the spectrum and of the eigenvectors of the coupling matrix $J$ during the dreaming procedure. Before dreaming, the spectrum of $J$ is of the Marchenko–Pastur type \cite{marcenkoDISTRIBUTIONEIGENVALUESSETS1967}, and the $N$-dimensional vector space is split between a degenerate $N-P$ dimensional eigenspace orthogonal to all the memories, and a $P$ dimensional space spanned by the memories, split in non-degenerate eigenspaces. Fig. \ref{fig:HU_ranked_spectrum} focuses on the evolution under dreaming of the ranked spectrum of $J$. The evolution of the ranked spectrum indicates that HU is targeting, and reducing, the largest eigenvalues of the coupling matrix, while all other eigenvalues are increased by a constant amount at every dream, maintaining a traceless coupling. This leads to a plateau on the high end of the ranked spectrum. In \cref{fig:HU_panini} 
we qualify the evolution of the eigenvectors $\vec{\zeta}$ of the coupling matrix $J$ as a function of the dreaming number. 
For each D, eigenvalues are ranked from 1 to N. For each rank, we measure the overlap $\omega( \vec{\zeta}(D), \vec{\zeta}(D-1))$ between the corresponding eigenvector at step $D$ and at step $D-1$.
Eigenvalues in the same rank at different dreaming steps are connected by a continuous line, colored with a color code connected to $\omega$. For clarity, only lines corresponding to overlaps larger than $0.1$ are shown. 
As the dreaming procedure unfolds, the majority of the eigenvectors does not change much (blue lines), and lines do not cross. This means that eigenvalues evolve continuously, while the corresponding eigenvectors barely change. The highest and lowest part of the ranked spectrum, on the other hand, show some crossing of lines, and low values of the overlaps (in red). This is due to the eigenvalues becoming almost equal, leading to an effectively degenerate eigenspace, corresponding to the plateau in \cref{fig:HU_ranked_spectrum}.

These observations suggest the following alternative algorithm.

\subsection{Eigenvector dreaming}
\begin{algorithm}[H]
    \caption{EVdreaming}
    \label{alg:EVdreaming}
    \begin{algorithmic} 
        \State Initialize $J$ using Hebb's rule \cref{eq:hebbs_learning_rule}
        \For{$D = 1$ to $D_{max}$}
            \State 1-Find an orthonormal basis of eigenvectors $\zeta^\mu$ of $J$.
            \State 2-Select the eigenvector $\zeta^{u_D}$ with the largest absolute eigenvalue. 
            \State 3-Update $J_{ij} \gets J_{ij} - \epsilon \zeta_i^{u_D} \zeta_j^{u_D}$. 
            \State 4-Reset diagonal terms to zero $J_{ii}\equiv 0$
        \EndFor
    \end{algorithmic}
\end{algorithm}
In this algorithm, the update of the couplings reduces the value of the highest eigenvalue by an amount $\epsilon$, leaving the eigenvectors unchanged. Resetting the diagonal to zero, on the other hand, increases the value of every eigenvalue by a stochastic amount (see \cref{subsec:IEVdreaming}), and also modifies the eigenvectors. Each step of this algorithm is based on the spectrum of the current coupling matrix. Note that this algorithm could be implemented using purely local rules, by iterating the synchronous update $\sigma^{t+1} = f (J\sigma^t)$ with$f(x) = \frac{x}{||x||_2}$, which converges towards the eigenvector of $J$ with the largest eigenvalue.
\subsection{Initial Eigenvector dreaming}
\label{subsec:IEVdreaming}
An even simpler dreaming procedure, which does reproduce the qualitative features of HU (specifically the centrality of the spectrum evolution and the marginality of the eigenspaces evolution) is obtained by modifying the coupling matrix on the basis of the eigenvectors of the \textit{initial} coupling matrix $J^{H}$, as listed in \cref{alg:IEVdreaming}. We call this procedure \textit{Initial Eigenvector dreaming} (IEVdreaming).
\begin{algorithm}[H]
    \caption{IEVdreaming}
    \label{alg:IEVdreaming}
    \begin{algorithmic} 
        \State 1-Initialize $J$ using Hebb's rule \cref{eq:hebbs_learning_rule}
        \State 2-Find an orthonormal basis of eigenvectors $\zeta^\mu$ of the initial coupling matrix.
            \For{$D = 1$ to $D_{max}$}
            \State 3-Consider the most recent coupling matrix $J^{D-1}$, and select the eigenvector $\zeta^{u_D}$ with the largest absolute eigenvalue. 
            \State 4-Update $J_{ij} \gets J_{ij} - \epsilon \zeta_i^{u_D} \zeta_j^{u_D}$. 
            \State 5-Remove the average value of the diagonal elements of $J$: $J_{ii} \gets J_{ii} - \frac{\epsilon}{N}$.
        \EndFor
    \end{algorithmic}
\end{algorithm}
This algorithm is simple enough that it can be analyzed in some detail.

\subsection{A first analysis of IEVdreaming}
\label{subsec:IEVdreaming_analysis}
\begin{figure}[t]
\centering
\includegraphics[width=1\linewidth]{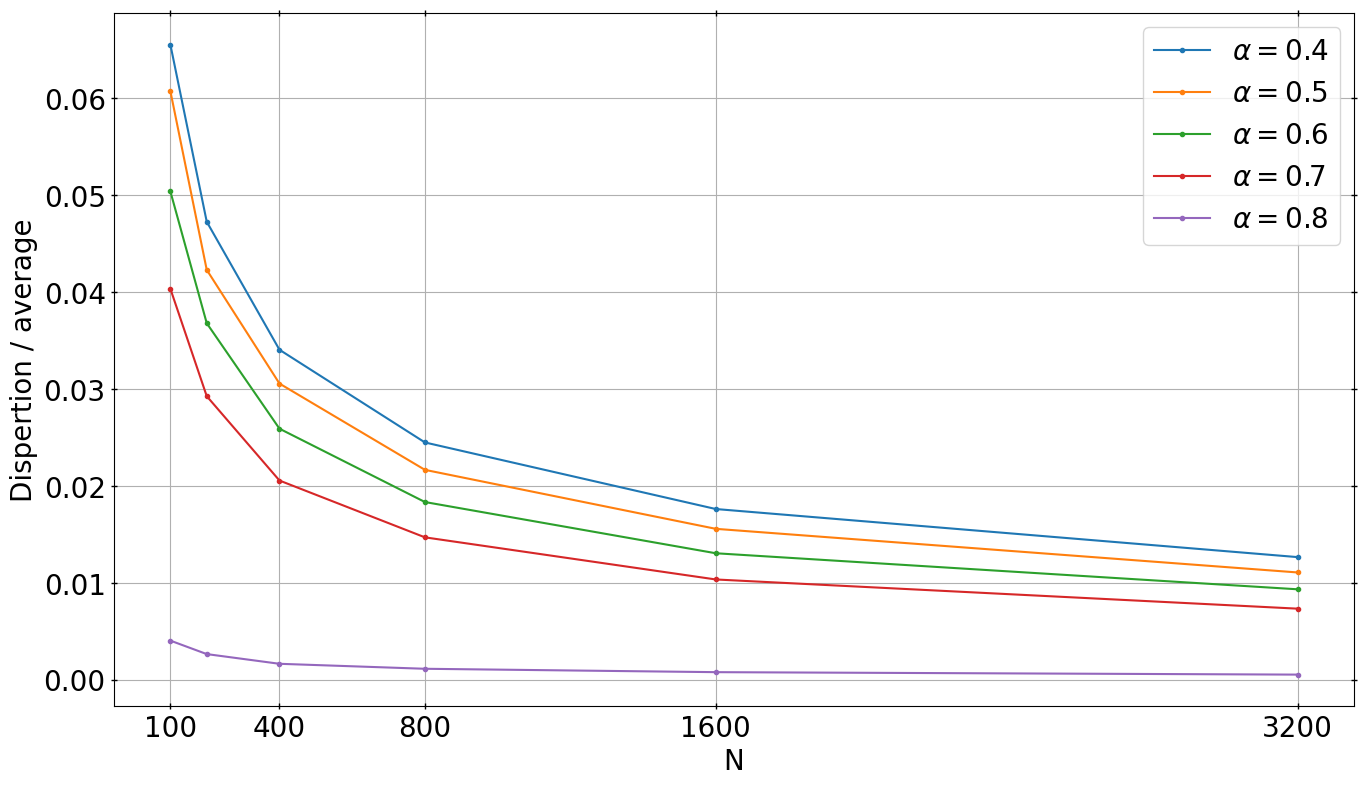}
\caption{Dispersion of the corrections to the low plateau eigenvalues, divided by the low plateau eigenvalue, at $D_{top}$, as a function of N, for different values of $\alpha$. As the system size is increased, the corrections become negligible compared to the low plateau eigenvalue.}
\label{fig:eta_is_negligible}
\end{figure}

As a first approach, imagine removing step 5 of the iterative process, and simply setting the diagonal to zero after the for cycle. The resulting J reads
\begin{equation}
        \begin{aligned}
        J_{ij}^D & = \sum_{\mu = 1}^N \zeta_i^\mu \zeta_j^\mu \left( \lambda_\mu - \epsilon \sum_{d=1}^D \delta_\mu^{u_d} \right) + \epsilon \sum_{d=1}^D (\zeta_i^{u_d})^2\delta_{ij}\\
        & = \sum_{\mu = 1}^N \zeta_i^\mu \zeta_j^\mu \left( \lambda_\mu - \epsilon \sum_{d=1}^D \delta_\mu^{u_d} \right) + 
        \epsilon \sum_{d=1}^D \langle (\zeta_i^{u_d})^2 \rangle\delta_{ij} + \\
        &\quad+\epsilon \sum_{d=1}^D \Big[ (\zeta_i^{u_d})^2- \langle (\zeta_i^{u_d})^2 \rangle\Big]\delta_{ij}\;,
        \end{aligned}
\end{equation}
where the average $\langle (\zeta_i^{u_d})^2 \rangle$ is computed over the statistics generated by the choice of the eigenvector $u_D$ to be dreamed at each step, given the realization of disorder (i.e. the value of the eigenvectors $\zeta_i^\mu$). Since the eigenvectors of a Wishart matrix are isotropically distributed on the $(N-1)\text{-dimensional}$ sphere, one has that $\langle (\zeta_i^{u_d})^2 \rangle=1/N$.
The result is then 
\begin{equation}
    J_{ij}^D  \simeq \sum_{\mu = 1}^N \zeta_i^\mu \zeta_j^\mu \left( \lambda_\mu - \epsilon \, d_\mu \right) + \epsilon \frac{D}{N}\delta_{ij} + \eta_{ij}\,,
    \label{eq:J_IEVec}
\end{equation}
where $d_\mu=\sum_{D=1}^D \delta_\mu^{u_D}$ and $\eta_{ij}$ is a diagonal random matrix
\begin{equation}
    \label{eq:eta_matrix}
    \eta_{ij}\equiv\epsilon \sum_{d=1}^D \Big[ (\zeta_i^{u_d})^2- \langle (\zeta_i^{u_d})^2 \rangle\Big]\delta_{ij}\;.
\end{equation}
The first two terms preserve the eigenvectors of $J$. The $\eta$ correction changes both the eigenvectors and eigenvalues of the coupling matrix, and  assuming that $\eta$ is small enough, we can compute those changes perturbatively. In particular, the degenerate eigenspace corresponding to the low eigenvalue plateau will be split by corrections $\lambda\to\lambda+\delta\lambda_i$, $i=1,...,N-P$ given by the $N-P$ eigenvalues of the matrix
\begin{equation}
    A^{\mu\nu}\equiv\zeta{^{\mu}}^\top \eta \zeta^\nu, \qquad\mu,\nu=1,...,N-P\,,
    \label{eq:correction_matrix}
\end{equation}
where the eigenvectors all belong to the low eigenvalue degenerate plateau (any orthonormal set of eigenvectors is equivalent).  In the thermodynamic limit, the impact of $\eta$ on $J$ becomes negligible, as shown in  \cref{fig:eta_is_negligible}. The $x$-axis represents $N$. The $y$-axis represents the eigenvalues of the $A$ matrix \cref{eq:correction_matrix} divided by the absolute height of the low plateau. In the thermodynamic limit, all curves tend to zero, showing that the corrections become negligible compared to the low plateau value. Some insight into this behavior can be gained by considering the statistics of the diagonal element of $\eta$. Their average is zero, by definition. If the $\xi^\mu_i$ involved in \cref{eq:eta_matrix} were a finite number, they could be treated as i.i.d. normal variables $\mathcal{N}(0,1/N)$, and the statistics of $\eta$ could be heuristically understood as proportional to a $\chi^2$ distribution, whose variance scales as $1/N$ (this is not exact, since not every eigenvector is dreamed the same number of times). Since we are dreaming an extensive number of eigenvectors, the $\xi^\mu_i$ are not independent (for one thing, they are constrained by normalization $\sum_{\mu=1}^N\xi^\mu_i=1$). Intuitively though, this has the effect of reducing the variance of $\eta_{ii}$. Hence, the $\chi^2$ distribution is an upper bound for the size of $\eta$, going to zero. Given this, the dreaming procedure is described by the simple update rule
\begin{equation}
    J_{ij}^D  \simeq \sum_{\mu = 1}^N \zeta_i^\mu \zeta_j^\mu \left( \lambda_\mu - \epsilon d_\mu \right) + \epsilon \frac{D}{N}\delta_{ij}.
    \label{eq:J_IEVec_analytic}
\end{equation}
This algorithm is very inexpensive from the computational point of view, since one does not need to compute eigenvectors multiple times.

Whether the correction to the diagonal elements of $J$ is carried out at each step of the algorithm or at the end, affects the choice of the eigenvector that gets dreamed: if the correction is carried out at the end, the negative degenerate plateau will quite soon be higher in absolute value than the high plateau (we call this \textit{inversion}). Then, the algorithm will start selecting eigenvectors from the low plateau, which are orthogonal to the memories, having no effect on the stabilities. On the other hand, the choice in \cref{alg:IEVdreaming} reproduces the qualitative behavior of HU in an analytically simple setting, since taking out the diagonal at each step decreases the absolute value of the low negative plateau while increasing the absolute value of the positive plateau, delaying the inversion.

\section{Algorithm performance}
\begin{figure}[t]
\centering
\includegraphics[width=1\linewidth]{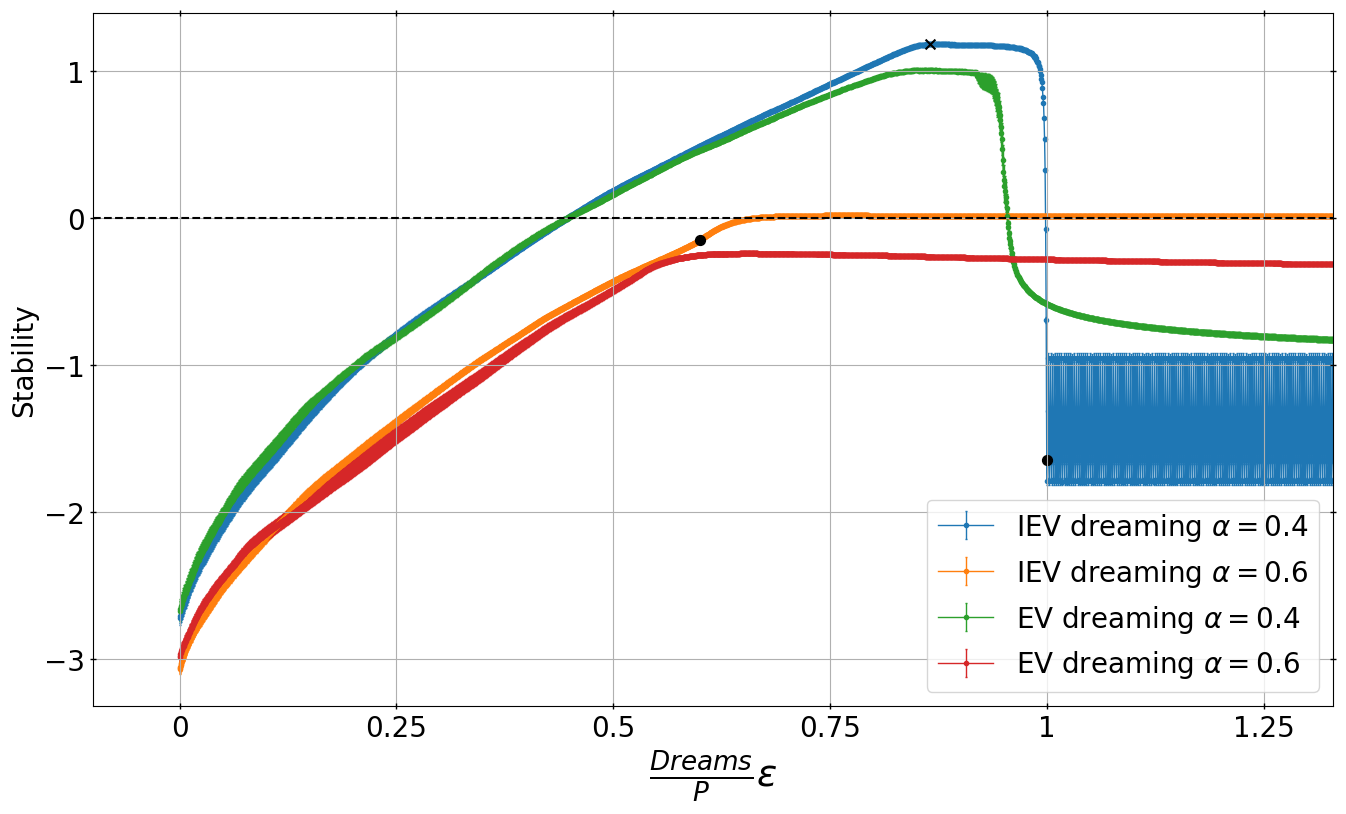}
\caption{Evolution of $\Delta_{min}$ while iterating different dreaming procedures, for some $\alpha$ values. $N=400$, $\epsilon=0.001$. $D_{top}$ is indicated by a cross, $D_{inv}$ is indicated by a dot. The new algorithms have very similar performances before $D_{inv}$, indicating the IEVdreaming is indeed a good model of EVdreaming.}
\label{fig:stab_all}
\end{figure}

In \cref{fig:stab_all} we show representative examples of the evolution of $\Delta_{min}$ according to the different dreaming procedures. The newly introduced algorithms have very similar performance before the inversion point $D_{inv}$ (marked by circles on the curves  in \cref{fig:stab_all}). This also indicates that the IEVdreaming is indeed a good model of EVdreaming. They also display the same qualitative behavior as HU.  In \cref{fig:stab_all}, crosses on the curves indicate when the algorithms start dreaming for the first time the lowest eigenvalue of the high portion of the ranked spectrum. This condition corresponds to the highest portion of the ranked spectrum becoming a plateau. In our new procedures this instant is very close to $D_{top}$. After $D_{top}$, IEV and EV display a plateau in the stability curve, which lasts until the inversion point, marked by dots in the curves. After the inversion point, which experimentally happens first in EVdreaming, EV and IEV display different behaviors, since the procedure becomes very sensitive to the eigenvectors dreamt. The behavior of IEV dreaming is detailed in \cref{sec:IEV_analytical}.

\begin{figure}[t]
\centering
\includegraphics[width=1\linewidth]{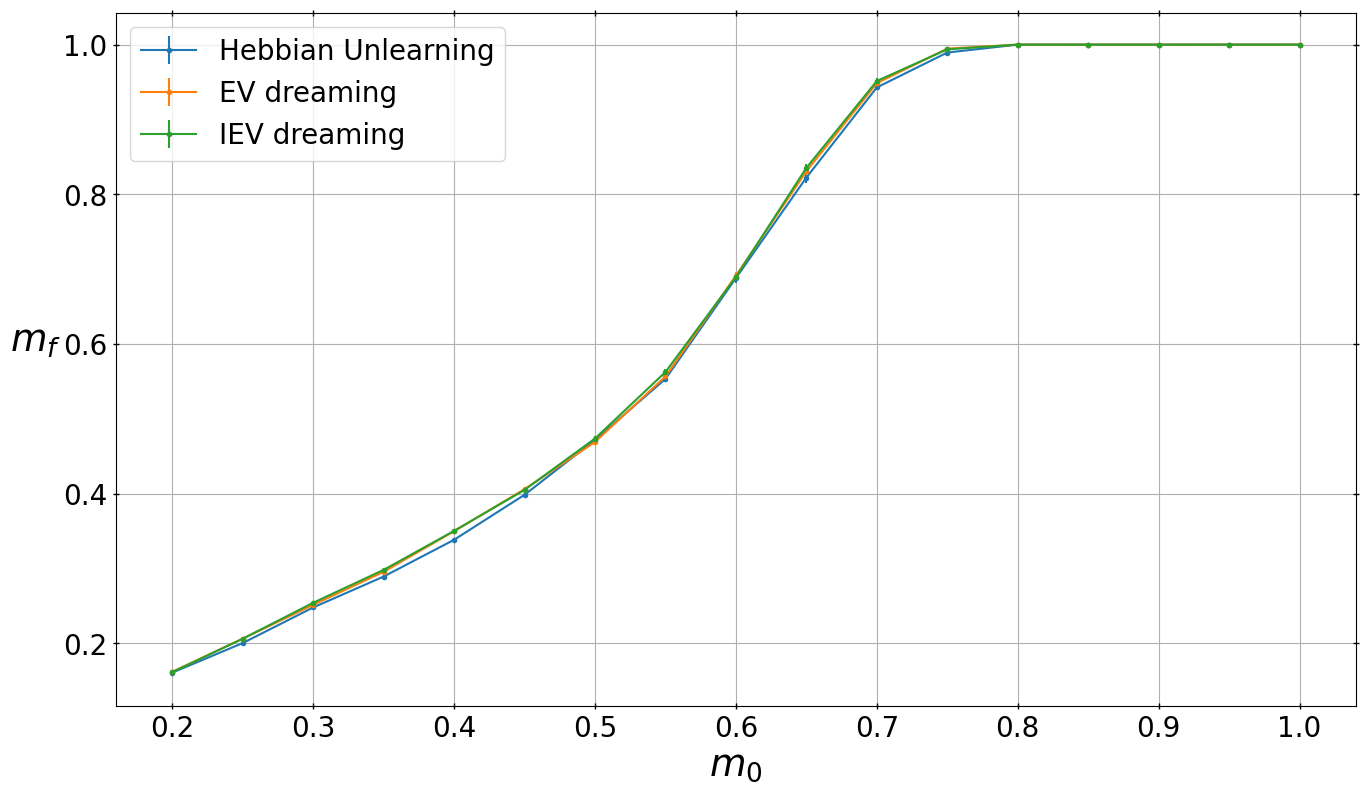}
\caption{Retrieval mapping for the various dreaming procedures, at $D=D_{in}$, where attraction basins are the largest. $N=400$, $\alpha=0.4$, $\epsilon=0.01$. Different curves coincide, suggesting that our new dreaming procedures capture the essence of HU.}
\label{fig:mf_mi_all}
\end{figure}

In \cref{fig:mf_mi_all} we compare the different algorithms in terms of the retrieval mapping, at $d=D_{in}$, where the performance is optimal. The quantitative differences in the $\Delta_{min}$ profile between the algorithms are reduced to virtually no difference, when the retrieval mapping is concerned.  Below the critical load wide basins of attractions are produced around the memories.

Defining the critical capacity of an algorithm $\alpha_c$ as the highest load such that $\Delta_{min} > 0$ is reached before $D_{inv}$, we find $\alpha_c^{IEVd}\sim 0.57$ and $\alpha_c^{EVd}\sim 0.55$, to be compared with $\alpha_c^{HU}\sim 0.59$.

\section{Analytical characterization of IEVdreaming}
\label{sec:IEV_analytical}
\begin{figure}[t]
\centering
\includegraphics[width=1\linewidth]{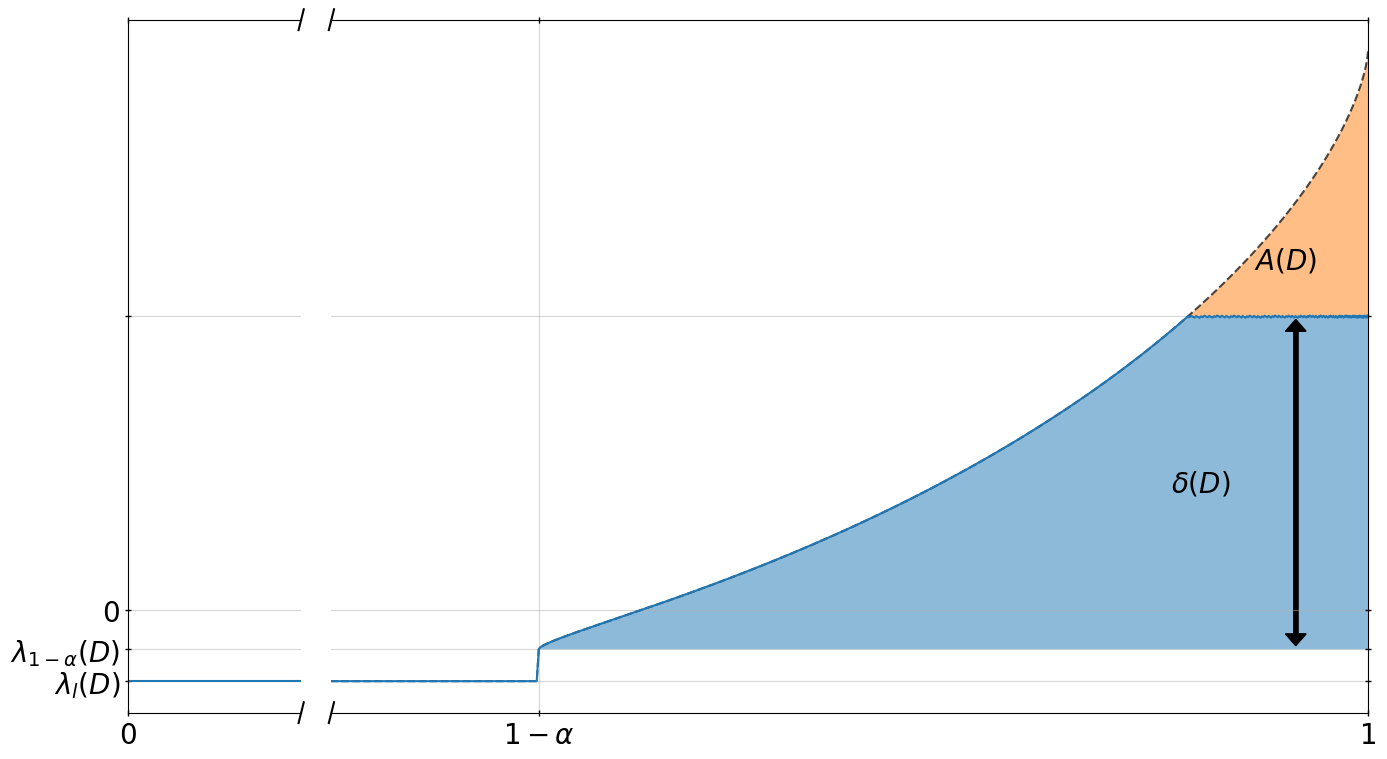}
\caption{Evolution of the ranked spectrum during IEVdreaming}
\label{fig:IEV_analytical}
\end{figure}

In the case of IEVdreaming, both the values of $D_{top}$ and $D_{inv}$ can be computed analytically. Let us define by $\lambda_l(D)$ the height of the low plateau, by $\lambda_{1-\alpha}(D)$ the height of the lowest eigenvalue in the high part of the ranked spectrum, and by $\delta(D)$ the distance between the high plateau and $\lambda_{1-\alpha}(D)$ (see \cref{fig:IEV_analytical}).
Before dreaming, one has 
\begin{align}
    &\lambda_l(0)=-\alpha\\
    &\lambda_{1-\alpha}(0)=1-2\sqrt{\alpha}\\[4pt]
    &\delta(0)=4\sqrt{\alpha}\;.
\end{align}
At each dream, the change in the ranked spectrum consists of an increase of every eigenvalue due to the resetting to zero of the diagonal elements of $J$, and a decrease of the dreamed eigenvalue, as per \cref{eq:J_IEVec_analytic}. Prior to $D_{top}$, i.e. before the high part of the ranked spectrum is completely flattened into a plateau, the evolution of the spectrum can be characterized by: 
\begin{align}
    &\lambda_l(D)=\lambda_l(0)+\frac{\epsilon D}{N}\\
    &\lambda_{1-\alpha}(D)=\lambda_{1-\alpha}(0)+\frac{\epsilon D}{N}\;,
\end{align}
while $\delta(D)$ can be determined numerically, noting that the area $A(D)$ is
\begin{equation}  
    A(D)=\frac{\epsilon D}{N}.
\end{equation}
Similar geometrical reasoning for $D>D_{top}$ leads to even simpler equations:
\begin{align}
    &\lambda_l(D)=\lambda_l(D_{top})+\frac{\epsilon (D_{top}-D)}{N} \label{eq:atd_Dtop_1}\\
    &\lambda_{1-\alpha}(D)=\lambda_{1-\alpha}(D_{top})+\frac{\epsilon (D_{top}-D)}{N}\Big(1-\frac{1}{\alpha}\Big)\label{eq:atd_Dtop_2}\\[9pt]
    &\delta(D)=0\;.
\end{align}
Given these relations, $D_{top}$ and $D_{inv}$ are determined by 
\begin{align}
    &\delta\big(D_{top}\big)=0 \\[9pt]
    &\big|\lambda_{l}(D_{inv})\big|=\big|\lambda_{1-\alpha}(D_{inv})+\delta(D_{inv})\big|\;.
\end{align}
These theoretical results for $D_{top}$ and $D_{inv}$ are compared to the results of the numerical simulations in \cref{fig:IEV_D_{inv}_theoVSsim}, with excellent agreement. 
\begin{figure}[t!]
\centering
\includegraphics[width=1\linewidth]{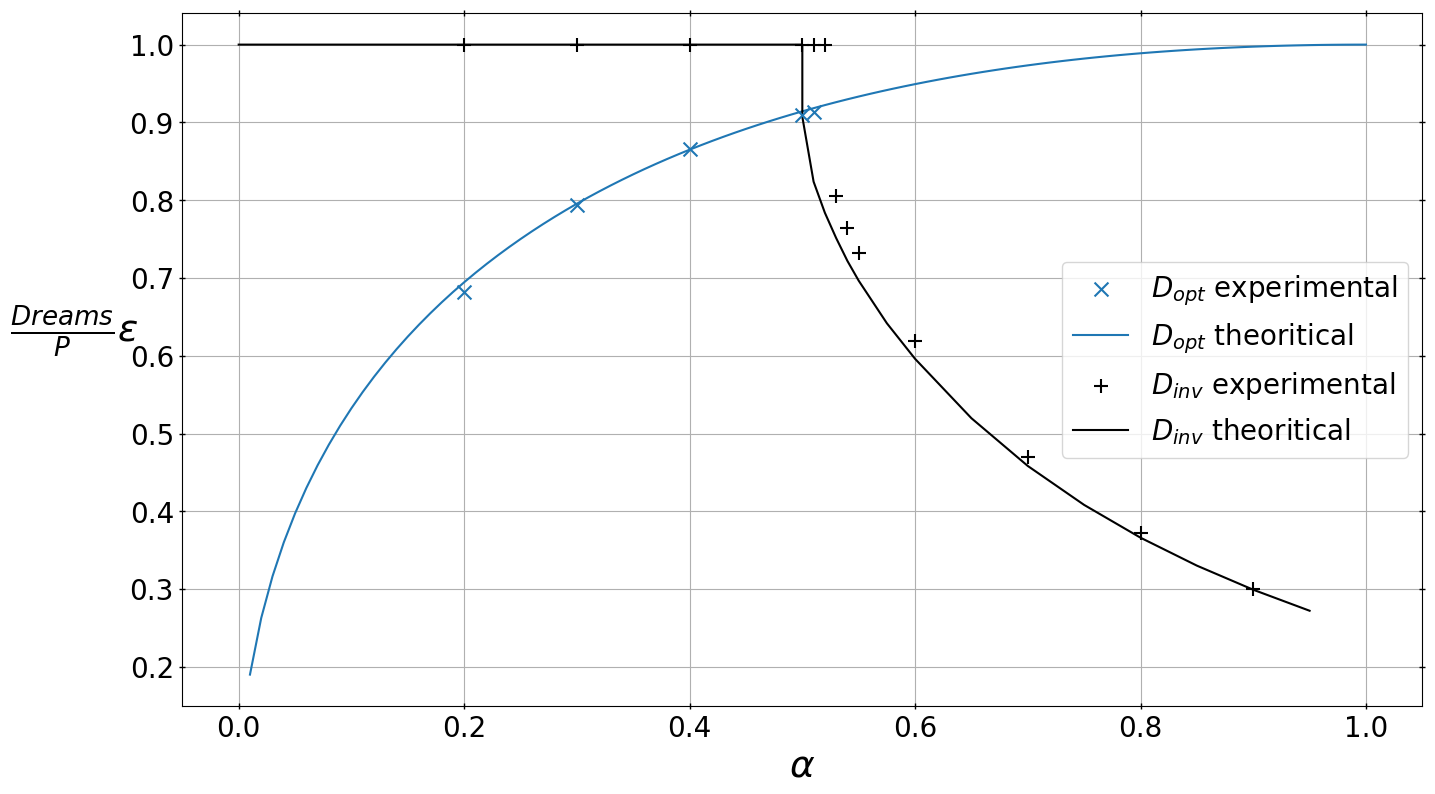}
\caption{Comparison between analytical estimate and simulations for $D_{inv}$ and $D_{top}$ as a function of $\alpha$. Parameters for the simulations are $N=1000$, $\epsilon=0.001$. The agreement is excellent, as finite size effects are already small at this size.}
\label{fig:IEV_D_{inv}_theoVSsim}
\end{figure}

In IEV dreaming, the evolution of the stabilities is determined exclusively by the evolution of the spectrum of $J$, since the eigenvectors do not change.
\begin{equation}
    \Delta_i^\mu=\xi_i^\mu \frac{\sum_{\nu=1}^N \lambda_\nu \zeta_i^\nu  w_\nu^\mu}{\sqrt{\sum_{\nu=1}^N\left(\lambda_\nu \zeta_i^\nu\right)^2}},
\end{equation}
where $w_\nu^\mu$ are the coordinates of the memories in the basis of the eigenvectors
\begin{equation}
    w_\nu^\mu\equiv(\boldsymbol{\zeta}^\nu \cdot\boldsymbol{\xi}^\mu)\;.
\end{equation}
After $D_{top}$, when the spectrum is composed by two plateaus $\mathcal{P}\pm$, this expression simplifies to
\begin{equation}
    \Delta_i^\mu = \xi_i^\mu  \frac{ \sum_{\nu \in \mathcal{P}_+}   \zeta_i^\nu  w_\nu^\mu}{\sqrt{ \sum_{\nu \in \mathcal{P}_+} \left(\zeta_i^\nu \right)^2 +  \left(\frac{\lambda_{l}(D)}{\lambda_{1-\alpha}(D)}\right)^2 \sum_{\nu \in \mathcal{P}_-}  \left( \zeta_i^\nu \right)^2 }}\;,
\end{equation}
which is constant (after $D_{top}$) as a consequence of \cref{eq:atd_Dtop_1,eq:atd_Dtop_2}. This explains the plateaus in \cref{fig:stab_all}.

For $\alpha<0.5$, one has $D_{inv}=P/\epsilon$, and $\lambda_l(D_{inv})=\lambda_{1-\alpha}(D_{inv})=0$. This means that at $D_{inv}$ we have $J=0$. In numerical simulations, given the finite value of $\epsilon$, this never happens. Instead, from $D_{inv}$ the network dreams every eigenvector of the high plateau, making it smaller than the low plateau, and then every eigenvector in the low plateau. Over $N$ dreams, all eigenvectors have been dreamed once. Thus, each eigenvalue is decreased once by $- \epsilon$ and increased $N$ times by $\frac{\epsilon}{N}$, restoring it to the initial value. This reflects in a periodic behavior of $\Delta_{min}$, which oscillates (see \cref{fig:stab_all}). For $\alpha>0.5$, on the other hand, the inversion happens with well separated plateaus $\lambda_l(D_{inv})<0<\lambda_{1-\alpha}(D_{inv})$. Hence, around $D_{inv}$, when the high plateau and the low plateau become closer than $\epsilon$ in absolute value, the network starts dreaming one eigenvector of the low plateau. At each dream, the corresponding eigenvalue is made even smaller, i.e. bigger in absolute value, and the network gets stuck dreaming it repeatedly. Asymptotically, this eigenvector (orthogonal to the memories) dominates the coupling matrix, leading again to zero stability without oscillations (see \cref{fig:stab_all}).

\section{Conclusions}
In this paper we unveiled an interesting feature of Hebbian Unlearning, namely the fact that eigenvectors of the coupling matrix do not change significantly during the algorithm, and the improvement in recognition performance is mostly due to a modification of the spectrum. Starting from this observation, we have proposed two new effective unlearning algorithms: Eigenvector dreaming and Initial Eigenvector dreaming, which emphasize the splitting of the learning problem into a trivial eigenvector evolution and a non-trivial spectrum evolution. IEVdreaming is the simplest algorithm, being computationally efficient and easy to control analytically. IEVdreaming turns out to give a very good description of EVdreaming, and a qualitatively good description of HU. Finally, in our new algorithms, we found a strong correlation between the moment when lowest eigenvalues of the high plateau starts being dreamed, and the moment when the algorithm stops increasing the minimum stability $\Delta_{min}$. This correlation, which follows from simple analytical arguments in the case of IEV dreaming, is also present, to a lesser extent, in HU.

\section{Acknowledgments}
EM acknowledges funding from the PRIN funding scheme
(2022LMHTET - Complexity, disorder and fluctuations: spin glass physics and beyond) and from the FIS (Fondo Italiano per la Scienza) funding scheme (FIS783 - SMaC - Statistical Mechanics and Complexity: theory meets experiments in spin glasses and neural networks) from Italian MUR (Ministery of University and Research). 
MM acknowledges financial support by the PNRR-PE-AI FAIR project funded by the NextGeneration EU program.

\end{document}